\documentclass[10pt,a4paper,twoside,english]{amsart}

\usepackage[all]{xy}
\CompileMatrices
\usepackage{amsmath,amssymb,latexsym}
\usepackage{graphics}

\def\e{\varepsilon}
\def\s{\sigma}
\def\bc{\begin{center}}
\def\ec{\end{center}}
\def\be{\begin{equation*}}
\def\ee{\end{equation*}}
\def\ba{\begin{array}}
\def\ea{\end{array}}
\def\bp{\begin{picture}}
\def\ep{\end{picture}}

\def\bea{\begin{eqnarray*}}
\def\eea{\end{eqnarray*}}
\def\bnea{\begin{eqnarray}}
\def\enea{\end{eqnarray}}

\def\lra{\longrightarrow}
\def\Lra{\Longrightarrow}

\def\lmt{\longmapsto}
\def\bc{\begin{center}}
\def\ec{\end{center}}
\def\noi{\noindent}

\def\AG{\mathrm{Aut} \, (G)}

\def\OG{\mathrm{Out} \, (G)}

\def\LG{\mathfrak{g}}
\def\LH{\mathfrak{t}}

\def\RG{{\mathbf{Rep}}(G)}

\def\A{\mathbf{A}}

\def\ad{\mathrm{ad}}

\def\p{\partial}
\def\c{\mathsf{c}}
\def\d{\mathsf{d}}

\def\de{\textbf{Definition} : }

\def\MP{\mathcal{P}}
\def\N{\mathbb{N}}
\def\R{\mathbb{R}}

\def\tr{\mathrm{tr}\,}
\def\Z{\mathbb{Z}}

\def\V{C_0}
\def\L{C_1}
\def\P{C_2}

\def\AT{\widetilde{A}}

\def\LT{\widetilde{C}_1}
\def\PT{\widetilde{C}_2}
\def\QT{\widetilde{C}_3}

\def\pic{\xymatrix}
\def\vs{\vspace{5mm}}
\def\nl{\newline}
\def\noi{\noindent}

\begin{document}

\fontfamily{cmss}
\fontseries{m}
\selectfont

\title{Homotopy and duality in non-Abelian lattice gauge theory}

\author{Romain Attal}

\email{attal@lpthe.jussieu.fr}

\address{Laboratoire de Physique Th\'{e}orique et 
Mod\'{e}lisation \nl
Universit\'e de Cergy-Pontoise \nl
5, Mail Gay-Lussac \nl
F-95031 Neuville-sur-Oise (FRANCE)}

\hfill\today
\vs

\begin{abstract}

We propose an approach of lattice gauge theory based on a homotopic interpretation 
of its degrees of freedom. The basic idea is to dress the plaquettes of the lattice 
to view them as elementary homotopies between nearby paths. Instead of using a unique 
$G$-valued field to discretize the connection 1-form, $A$, we use an $\AG$-valued 
field $U$ on the edges, which plays the role of the 1-form $\ad_A$, and a $G$-valued 
field $V$ on the plaquettes, which corresponds to the Faraday tensor, $F$. 
The 1-connection, $U$, and the 2-connection, $V$, are then supposed to have a 
2-curvature which vanishes. This constraint determines $V$ as a function of $U$ up to 
a phase in $Z(G)$, the center of $G$. The 3-curvature around a cube is then Abelian 
and is interpreted as the magnetic charge contained inside this cube. 
Promoting the plaquettes to elementary homotopies induces a chiral splitting of their 
usual Boltzmann weight, $w=v\bar{v}$, defined with the Wilson action. We compute 
the Fourier transform, $\widehat{v}$, of this chiral Boltzmann weight on $G=SU_3$ 
and we obtain a finite sum of generalized hypergeometric functions. The dual model 
describes the dynamics of three spin fields : 
$\lambda_P\in{\widehat{G}}$ and $m_P\in{\widehat{Z(G)}}\simeq\Z_3$, 
on each oriented plaquette $P$, and $\e_{ab}\in{\widehat{\OG}}\simeq\Z_2$,
on each oriented edge $(ab)$. Finally, we sketch a geometric interpretation of this 
spin system in a fibered category modeled on the category of representations of $G$.

\end{abstract}

\maketitle

\tableofcontents

\vs
\section{Introduction}
\vs

A discretization of Yang-Mills gauge theory on a lattice has been 
defined by K. Wilson in order to find an analytic criterion 
for the confinement of quarks \cite{W}. 
This approach suggests that confinement is a consequence of 
the non-Abelian nature of the structure group (usually $G=SU_N$). 
The gauge invariant quantities, i.e. the physical observables of the theory, 
are obtained by averaging products of traces of holonomies of a connection 
on a vector bundle, along spin networks. The latter are directed 
graphs whose edges carry a representation of $G$ and whose
vertices carry a homomorphism from the tensor product of the
incoming representations to the tensor product of the outgoing
ones. These observables can be expanded as power series whose
coefficients are sums over surfaces made of plaquettes \cite{W,DZ,MM}. 
In this framework, the correspondence between weak and strong 
coupling is expressed by a Fourier transformation on the space 
of gauge fields and is a non-Abelian generalization of the 
Kramers-Wannier duality \cite{KW}. In \cite{OP}, the dual theory 
has been formulated as a spin-foam model on the dual lattice. 
In the present work, we propose to make a further step 
in the study of non-Abelian duality by using homotopic methods.

Our paper is organized as follows. In Section 2, we define the
dressed cells of our lattice in order to view them as elementary
homotopies. They can be seen as the generators of the spaces of 
iterated paths.
In Section 3, we define the gauge fields of our system and 
their partition function, in which the 2-curvature is forced to vanish.
Since Wilson's action is proportional to the real part of the trace 
(in the fundamental representation of $G$) of the holonomy along a little
square loop, the Boltzmann weight, $w$, of an unoriented plaquette 
is the product of two complex conjugate weights, $v$ and
$\overline{v}$, each one being associated to an orientation of 
this plaquette. This chiral splitting allows to associate independant 
representations to the 2-cells of the dual lattice which correspond 
to oppositely oriented plaquettes. 
In Section 4, we express the partition function of the dual model in terms 
of three fields of representations (or spin fields) taking their values 
in ${\widehat{G}}$, ${\widehat{Z(G)}}$ and ${\widehat{\OG}}$.
Finally, in Section 5, we draw some conclusions and future prospects.

In the appendices, we collect some mathematical results.
In App. A, we recall some useful results about the structure of $SU_3$ and 
its characters.
In App. B, we give a dual interpretation of Wilson's action by proving
that $\widehat{w}$ satisfies a diffusion equation on the weight lattice of $SU_3$. 
In a Minkowskian space-time, this corresponds to the motion of 
a free particle following a discretized Schr\"{o}dinger's equation. 
The application of the spectral theorem to the corresponding discrete 
Laplacian provides us with a direct proof of Fourier's inversion
formula. 
In App. C, we compute explicitely the Fourier coefficients
of the chiral Boltzmann weight $v$ : this is a linear combination of 
generalized hypergeometric series.

\vs
\section{Homotopies on a lattice}
\vs

\subsection{From plaquettes to homotopy generators} 

The building blocks of a gauge theory on a lattice are the cells 
of this lattice. At the beginning, we consider them as mere subsets 
of the space in which the lattice is embedded. So, let $\Z^4$ 
be embedded in the four dimensional euclidian space $\R^4$, 
endowed with its canonical basis $(e_i)_{1 \leq i \leq 4}$. 
We will use the following notations :

\bea
A &=& \{ 1,2,3,4 \} \\
\AT &=& A \cup \{ 0 \} \cup (-A) \\
A_p &=& \{ i=(i_1,\cdots,i_p)\in A^p\ :\ i_1<\cdots<i_p\} \\
e_{-i} &=& - e_i \qquad\forall\,i\in\AT \\
u &=& (e_1+e_2+e_3+e_4)/2 \\
\eea

\vs
\de \emph{For each $p \in \{0,1,2,3,4\}$, $C_p$ denotes 
the set of bare (unoriented) $p$-cells}

\bea
C_p := \Z^4 \times A_p  \\
\eea
If we choose a characteristic map $(f_\s : B^p \to \R^4)$ to 
parametrize homeomorphically each $p$-cell $\s$ by the unit ball
$B^p\subset\R^p$, then we obtain a cell complex 
$C=(C_p)_{1\leq p\leq 4}$. 
The geometric realization of $\s\in C_p$ is the image of $f_\s$, 
noted $\vert\s\vert$, and the $p$-skeleton of $C$ is the sub-complex 
$C^p=(C_q)_{q\leq p}$. The terms "vertex", "link", "plaquette", 
"cube" and "hypercube" will denote the sets 
$\vert\s\vert\subset\R^4$, for $\s\in C_0, C_1, C_2, C_3$ 
and $C_4$, respectively.
In order to take into account the two possible orientations of 
each cell, we adjoin a sign and we define the sets of oriented
cells by

\bea
{C'}_p := C_p \times \{ -1,+1\} \\
\eea

\vs
\de \emph{The dual of $C'$ is the complex 
$C'^\ast=({C'}^\ast_p)_{0\leq p \leq 4}$ defined by}

\bea
{C'}^\ast_p := (\Z^4 + u) \times A_p \times \{-1,+1\} \\ 
\eea
We have a duality map

\bea
\ast : C' & \lra & {C'}^\ast \\
\s = (a,i) & \lmt & \s^\ast = (a+u-e_{i^\ast},i^\ast) \\
\eea
where $i^\ast=(i^\ast_1, \cdots , i^\ast_{4-p})$ 
is the ordered multi-index defined by

\bea
\{i^\ast_1, \cdots , i^\ast_{4-p} \} 
= A \backslash \{i_1, \cdots , i_p \} \qquad {\mathrm{and}}
\qquad i^\ast_1 < \cdots < i^\ast_{4-p} \\
\eea
and $e_{i^\ast}=(e_{i_1^\ast}+\cdots+e_{i_p^\ast})/2$.

When we work with a non-Abelian structure group, we need to 
dress the bare cells with a supplementary structure in order 
to interpret them as elementary homotopies.

\vs
\de \emph{The set $\widetilde{C}_p$ of homotopic $p$-cells is defined by}

\bea
\widetilde{C}_p := {C'}_0 \times \AT^p  \\
\eea
Therefore, we have the inclusions

\bea
C_p \subset {C'}_p \subset \widetilde{C}_p \\
\eea
A homotopic 0-cell is simply a vertex with a sign.
A homotopic 1-cell $\ell=(a,i,\e)\in\LT$ 
has a 0-source and a 0-target defined by

\bea
&& s_0 (a,i,1) = t_0 (a,i,-1) = a \\
&& t_0 (a,i,1) = s_0 (a,i,-1) = a+e_i \\
\eea

The fluxes associated to the various ways of sweeping a given plaquette
are related by a representation of the symmetry group of the square. 
Indeed, the 2-cells which start at $\ell$ are of the form $(\ell,j,1)$ with 
$j \in \AT$ and the 2-cells which contain $\ell$ are obtained by 
the action of the following group :

\bea
E &:=& D_4 \times D_1 \\
D_4 &=& \langle r,s \, \vert \, r^4=s^2=(rs)^2=1 \rangle \\
D_1 &=& \langle \iota \, \vert \, \iota^2=1 \rangle \\
\eea
$D_4$ acts on $\{a,b,c,d\}$ by

\bea
s(a)=d & \qquad & s(b)=c \\
s(c)=b & \qquad & s(d)=a \\
r(a)=d & \qquad & r(b)=a \\
r(c)=b & \qquad & r(d)=c \\
\eea
and this induces an action of $E$ on $\PT$ :

\bea
r(x,i,j,\e) &=& (x+e_i,j,-i,\e) \\
s(x,i,j,\e) &=& (x+e_i,-i,j,\e) \\
\iota (x,i,j,\e) &=& (x,i,j,-\e) \\
\eea
Therefore, the subset of non-degenerate homotopic 
2-cells is a $D_4$-bundle over $\P$.

The conjugation map on $\widetilde{C}_2$
can also be extended to $\widetilde{C}$ :

\bea
\widetilde{C}_p & \lra & \widetilde{C}_p \\
\tau=(x,j,\e) & \lmt & 
{\overline{\tau}}=(x,j,-\e) \\ 
\eea

\vs
\subsection{Iterated paths}

The loop approach of lattice gauge theories \cite{W,P,GP}
suggests to work in the spaces of iterated paths of our lattice. 
We propose the definition given below, but a more careful treatment
should use the notion of cellular homotopy $n$-groupoid.

\vs
\de \emph{The set of paths is defined by}

\bea
\MP_1 := \{ (a_0, \cdots , a_p) \ : \ a_i \in \Z^4 \quad
{\mathrm{and}} \quad \Vert a_{i+1} -a_i \Vert = 1 
\quad {\mathrm{for}} \quad i=0, \cdots, p-1 \} \\
\eea
Let $\gamma=(a_0 , \cdots , a_p)$ and 
$\gamma'=(a'_0,\cdots , a'_q)$ be two paths.
If $t_0(\gamma)=s_0(\gamma')$, i.e. $a_p=a'_0$, 
$\gamma$ and $\gamma'$ can be composed by concatenation :

\bea
\gamma'\cdot\gamma = (a_0, \cdots, a_p, a'_1, \cdots, a'_q) \\
\eea
The composition of $\gamma$ with the reversed family,
$(a_p,\cdots,a_0)$, provides a path which is homotopic to 
$(a_0,a_0)$ in $\vert C^1\vert$. 
Similarly, the 0-source and 0-target of a 2-cell 
$P=(a,i_1,i_2,\e)\in{\widetilde{C}}_2$ are the vertices defined by

\bea
&& s_0(P) = a \\
&& t_0(P) = a+e_{i_1} \\
\eea
and its 1-source and 1-target are the paths defined by

\bea
&& s_1(a,i_1,i_2,+1) = s_1(a,i_1,i_2,-1) = (a,a+e_{i_1}) \\
&& t_1(a,i_1,i_2,+1) = s_1(a,i_1,i_2,-1) = 
(a,a+e_{i_2},a+e_{i_2}+e_{i_1},a+e_{i_1}) \\
\eea
We will write generically

\bea
(a,b,c,d)=(a,a+e_j,a+e_j+e_i,a+e_i) \\
\eea
the four corners of $P=(a,i,j,\e)$ :

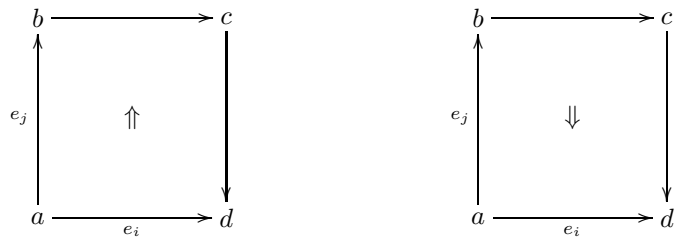
\vs
\[
\pic{
b \ar[rr] & & c \ar[dd] & & & b \ar[rr] & & c \ar[dd] \\
& \Uparrow & & & & & \Downarrow & \\
a \ar[uu]^{e_j} \ar[rr]_{e_i} & & d & & & a 
\ar[uu]^{e_j} \ar[rr]_{e_i} & & d\\
}
\]
\vs

\noi Roughly speaking, if $\e=+1$, $P$ represents a $(1 \to 3)$ 
homotopy and if $\e=-1$, $P$ represents a $(3 \to 1)$ homotopy.
We could use also $(0 \to 4)$ or $(2 \to 2)$ homotopies, i.e. 
diagonal motions, but the set $\PT$ defined above is sufficient 
to connect all the paths and respects the symmetry of the lattice 
by moving the paths only along the lattice vectors $(e_i)_{i \in B}$.
We allow degenerate 2-cells, with $i_2=\pm i_1$, which represent 
tangential motions of the links. Their boundary is homotopic, 
in $\vert C^1\vert$, to a constant path. 
Therefore, they carry no flux and do not contribute to the dynamics.

With higher dimensional cells, we can also define the sets 
$\MP_n$ of $n$-paths as follows. 

\vs
\de \emph{For $n>1$, a $n$-path is a finite family $(X_0, \cdots , X_p)$ 
of $(n-1)$-paths in which the composition of any two successive
members is homotopic, in $\vert C^{n-1}\vert$, to the boundary of a 
$(n-1)$-cell : }

\bea
\MP_n := \{ (X_0, \cdots , X_p) \ : \ X_i \in \MP_{n-1} \quad
{\mathrm{and}} \quad \exists \, \s\in C_n \quad {\mathrm{s. t.}} 
\quad X_{i+1}\circ X_i^{-1} \sim_{n-1} \p\s \}  \\
\eea
If two $k$-paths, say $\Gamma ,\Gamma' \in\MP_k$, are homotopic 
in the $k$-skeleton of $C$, i.e. if they can be joined by a sequence
of motions along degenerate $(k+1)$-cells, we will write
$\Gamma\sim_k\Gamma'$.

\vs

\de \emph{For $q<p$, we define the $q$-source and the $q$-target 
of a $p$-path in the obvious way, so that we have maps}

\bea
s_{qp} : \MP_p \lra \MP_q \\
t_{qp} : \MP_p \lra \MP_q  \\
\eea
which satisfy 

\bea
s_{rq} \circ s_{qp} &=& s_{rq} \circ t_{qp} \\
t_{rq} \circ s_{qp} &=& t_{rq} \circ t_{qp} \\
\eea

\vs
\section{Gauge fields and curvatures}
\vs

In the standard approach, the connection is represented 
(locally) by a map $(g:\V\times \AT \to G)$ from the set 
of 1-cells to the structure group, satisfying
$g_{\bar{\ell}}=g_\ell^{-1}$. 
Then, to each oriented plaquette with a base point, is
associated the holonomy of $g$ along its boundary. 
However, in the homotopic context, we need to separate
the degrees of freedom associated to the 1-cells from those 
associated to the 2-cells. This leads us to use two fields 

\bea
U : \LT & \lmt & \AG \\
V : \PT & \lmt & G \\
\eea
which will be called, respectively, the 1-connection and 
the 2-connection. We will also write

\bea
V_{abcd} &=& V_{(a,i,j,1)} \\
V_{\overline{abcd}} &=& V_{(a,i,j,-1)} \\
\eea
These fields are supposed to satisfy the following conditions :

\bea
U_{ba} = U_{ab}^{-1} & {\mathrm{and}} & V_{\overline{abcd}} = V_{abcd}^{-1} \\
V_{dabc} = U_{da} (V_{abcd}) & {\mathrm{and}} & V_{dcba} = U_{da} \big( \overline{V_{abcd}} \big) \\
\eea
which ensure that we have a natural representation of the space of paths. 
Indeed, the first two conditions mean that the motion along 
$ba$ (resp. $\overline{abcd}$) is represented by the inverse 
of the group element corresponding to $ab$ (resp. $abcd$). 
The third relation means that a rotation of angle $\pi /2$ applied 
to $abcd$ replaces the flux $V_{abcd}\in G$ by an equivalent flux 
$V_{dabc}$ obtained by applying the automorphism $U_{da}$, which
corresponds to the motion of the 0-target along the 1-source of
$abcd$. The second relation in (2) means that the flux swept by 
the flipped 2-cell $dcba = s(abcd)$ is the complex conjugate of 
the flux swept by $abcd$, pulled back to $s_0(s(abcd))=t_0(abcd)=d$.
Thus, we have only one independent variable $V$ per plaquette, 
the others being obtained by the action of $U$ or by inversion.

With the 1-connection and the 2-connection, we can define 
two kinds of curvature : the 2-curvature, defined on the 2-cells,
and the 3-curvature, defined on the cubes.

\vs
\de \emph{The 2-curvature $\Phi$ is the map defined by}

\bea
\Phi : \PT & \lra & \AG \\
abcd & \lmt & \Phi_{abcd} = I_{V_{abcd}} U_{ab}U_{bc}U_{cd}U_{da} \\
{\overline{abcd}} & \lmt & \Phi_{\overline{abcd}} = 
\Phi_{abcd}^{-1} \\
\eea
where $I$ denotes the inner automorphism map

\bea
I:G & \lra & \AG \\
g & \lmt & (I_g : h \mapsto ghg^{-1}) \\
\eea

\vs
\noi When the lattice spacing goes to zero, $\Phi$ corresponds to the 
"fake curvature" 2-form $\nu$ as defined in \cite{BM}.
Since $r^4(abcd)=abcd$ and since we want to represent the homotopy
along $abcd$ as a flux which depends only on the direction of
sweeping, we impose the relation

\bea
V_{abcd}=V_{r^4 (abcd)}=U_{ab}U_{bc}U_{cd}U_{da} (V_{abcd}) \\
\eea
which implies the vanishing of the 2-curvature :

\bnea
\boxed{
\Phi_{abcd} := I_{V_{abcd}} U_{ab}U_{bc}U_{cd}U_{da} 
= 1_{\AG} \qquad \forall \, abcd \in \PT
} 
\enea
\vs

\noi In order to describe the Yang-Mills-Wilson theory, we must impose 
another relation between $U$ and $V$, which is a discrete version 
of Bianchi's identity. Let $Q=(a,i_1,i_2,i_3,\e) \in \QT$ 
be a generic 3-cell : 

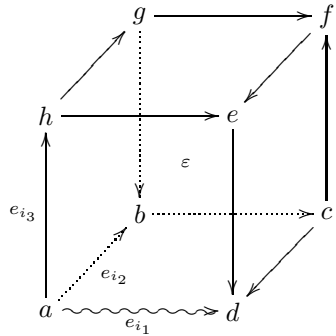
\vs
\[
\pic{
& g \ar@{.>}[dd] \ar[rr] \ar@{}[dddr]|-{\e} & & f \ar[dl] \\
h \ar[ur] \ar@{}[ddrr] \ar[rr] & & e \ar[dd] & \\
& b \ar@{.>}[rr] & & c \ar[uu] \ar[dl] \\
a \ar[uu]^{e_{i_3}} \ar@{.>}[ur]_{e_{i_2}} \ar@{~>}[rr]_{e_{i_1}} 
& & d & \\
}
\]
\vs

\noi Starting from the path $(ad)=s_1(Q)$, if we want to reach 
the path $(abcd)=t_1(Q)$, we can sweep either the 2-cell
$(ad,abcd)=s_2(Q)$ or the five remaining plaquettes, which 
form $t_2(Q)$ with the following sweeping scheme :

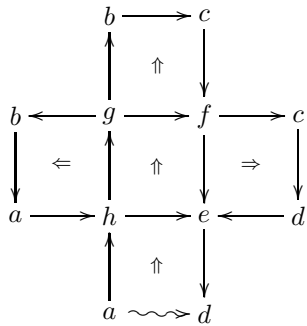
\vs
\[
\pic{
& b \ar[r] & c \ar[d] & \\
b \ar[d] & g \ar@{}[ur]|-{\Uparrow} \ar[u] \ar[l] \ar[r] 
& f \ar[r] \ar[d] & c \ar[d] \\
a \ar[r] & h \ar@{}[ul]|-{\Leftarrow} \ar@{}[ur]|-{\Uparrow} 
\ar[r] \ar[u] & 
e \ar@{}[ur]|-{\Rightarrow} \ar[d] & d \ar[l] \\
& a \ar@{}[ur]|-{\Uparrow} \ar[u] \ar@{~>}[r] & d & \\
}
\]
\vs

\noi Now we want to represent in $G$ the homotopy leading from 
$s_2(Q)$ to $t_2(Q)$. For this purpose, we make the product of 
the six fluxes (pulled back to $s_0(Q)=a$) which correspond to 
the six plaquettes swept in the order defined above and, 
if $\e=+1$, we put

\bea
\Omega_Q := 
U_{ad}(V_{defc})
\ V_{ahed} 
\ U_{ah}(V_{hgfe})
\ U_{ahg}(V_{gbcf})
\ V_{abgh}
\ V_{adcb} \\
\eea

\vs
\de \emph{The 3-curvature is the map}

\bea
\Omega : \QT & \lra & G \\
Q & \lmt & \Omega_Q \\
{\overline{Q}} & \lmt & \Omega_{\overline{Q}} = \Omega_Q^{-1} \\
\eea

\vs
{\textbf{Remark :}} Bianchi's identity can be stated as 
the vanishing condition for the 3-curvature $(\Omega_Q = 1_G)$. 
However, we don't impose this condition in our present model, 
because $\Omega_Q$ is the magnetic charge inside $Q$ and 
these topological defects could play a crucial role in the 
mechanism of quark confinement. In fact, Bianchi's identity is almost
satisfied thanks to the following elementary lemma.

\vs

{\textbf{Lemma : }} 
\emph{If the 2-curvature vanishes then the 3-curvature is abelian} :

\bea
\big( \Phi_P=1_{\AG}\quad\forall\, P\in\PT \big) & \Lra &
\big( \Omega_Q\in Z(G)\quad\forall\, Q\in\QT \big) \\
\eea
Let $Q\in\QT$. After the definition of $\Omega_Q$, we have 

\bea
I_{\Omega_Q} &=& 
\ I_{U_{ad}(V_{defc})}
\ I_{V_{ahed}}
\ I_{U_{ah}(V_{hgfe})}
\ I_{U_{ahg}(V_{gbcf})}
\ I_{V_{abgh}} 
\ I_{V_{adcb}} \\
&=& 
(U_{ad}I_{V_{defc}}U_{da})
\ I_{V_{ahed}}
\ (U_{ah}I_{V_{hgfe}}U_{ha})
\ (U_{ahg}I_{V_{gbcf}}U_{gha})
\ I_{V_{abgh}}
\ I_{V_{adcb}} \\
\eea
Using the relation (3), we obtain

\bea
I_{\Omega_Q} &=& 
(U_{ad}U_{dc}U_{cf}U_{fe}U_{ed}U_{da})
\ (U_{ad}U_{de}U_{eh}U_{ha})
\ (U_{ah}U_{he}U_{ef}U_{fg}U_{gh}U_{ha}) \\
&\quad & (U_{ahg}U_{gf}U_{fc}U_{cb}U_{bg}U_{gha}) 
\ (U_{ah}U_{hg}U_{gb}U_{ba})
\ (U_{ab}U_{bc}U_{cd}U_{da}) \\
&=& 1_{\AG} \\
\eea
Therefore, $\Omega_Q$ is in the kernel of $I$, 
i.e. in the center of $G$. \hfill $\Box$
\vs

\vs
\section{Partition function and duality}
\vs

We are now ready to write down the partition function of 
this lattice gauge theory defined in terms of homotopies.
Let $\tr$ denote the trace in the fundamental representation of $G$.
Let $\alpha = (3g^2T)^{-1}$, where $g$ is the Yang-Mills coupling and
$T$ denotes the temperature of the four dimensional system. 
The usual Boltzmann weight

\bea
w (M) = \exp \big( 2 \alpha \, 
{\mathrm{Re}}\, (\tr\, (M-1_G) ) \big) \\
\eea
is associated to an unoriented plaquette. However, since the elementary homotopies
along 2-cells carry an orientation, we will use as Boltzmann weight the function 
$(v : G \to \mathbb{C})$ defined by

\bea
v(M) := \exp \big( \alpha \, \tr \big( M-1_G \big) \big) \\
\eea
We thus have $w=\vert v \vert^2$. Let $N \in \N$ and let $\Lambda_N$ be the box defined by

\bea
\Lambda_N := \{ x \in \R^4 \, : \, \vert x_i \vert \leq N
\quad \forall \, i \in A \} \\
\eea
The degrees of freedom contained in this box are described 
by the maps $U$ and $V$ which satisfy the relations 
(1) to (3) and which are fixed to unity outside of $\Lambda_N$.
These maps form compact groups on which we have the 
product Haar measures, denoted $DU$ and $DV$.
The partition function of this system is then defined by

\bea
\mathcal{Z} := \int DU \int DV 
\prod_{P\in\P'} \delta (\Phi_P) \, v(V_P) \\
\eea
where $\delta (\Phi_P)$ denotes the Dirac distribution at
$1_{\AG}$. Note that the relation $w=v\overline{v}$ allows us 
to split the Boltzmann weight as

\bea
\prod_{P\in\P}w(V_P)=\prod_{P\in\P'}v(V_P) \\
\eea

\vs
\subsection{Duality transformation}

Let's apply Fourier's inversion formula 

\bea
v(M) = \sum_{\lambda \in \widehat{G}} d_\lambda \,
\widehat{v}_\lambda \, \overline{\chi}_\lambda (M) \\
\eea
to express $\mathcal{Z}$ as the partition function of a spin system : 

\bea
\mathcal{Z} &=& \int DU \int DV \, 
\bigg(
\prod_{P\in\P'}
\delta (\Phi_P) \, 
\sum_{\lambda \in \widehat{G}}
d_\lambda \,
\widehat{v}_\lambda \, 
\overline{\chi }_\lambda (V_P)
\bigg) \\
&=&\sum_\lambda
\bigg(
\prod_{P\in\P'} 
d_{\lambda_P} \, 
\widehat{v}_{\lambda_P} 
\bigg) 
\bigg( 
\int DU \int DV 
\prod_{P\in\P'} \, 
\delta (\Phi_P) \, 
\overline{\chi }_{\lambda_P} (V_P)
\bigg) \\
\eea
$\widehat{v}$ is computed in App. C. $\lambda$ now denotes a map 
from the set of oriented 2-cells to $\widehat{G}$ :

\bea
\lambda : \P' &\lra & \widehat{G} \\
P &\lmt & \lambda_P \\
\eea
so that  $\lambda_{abcd}$ and $\lambda_{dcba}$ are independent degrees of freedom. 
The constraint of vanishing 2-curvature can be solved as follows. 
In App. D, we show how $\AG$ can be described as a semi-direct product

\bea
\AG\simeq(G/Z)\rtimes\OG \\
\eea
This isomorphism is canonical and tells us that, for each edge $ab$, 
$U_{ab}$ defines a unique outer automorphism $\c^{\e_{ab}}$, where 
$\c\in\OG$ denotes the complex conjugation and $\e_{ab}\in \Z_2=\{0,1\}$.
Moreover, we can choose $T_{ab}\in G$ such that 

\bea
U_{ab} = \big( ZT_{ab}, \c^{\e_{ab}} \big) \\
\eea
where $Zg$ is the class of $g\in G$ in the coset $G/Z$. 
In terms of $T$ and $\e$, the 2-curvature becomes

\bea
\Phi_{abcd} = \big( ZV_{abcd}T_{ab}T_{bc}T_{cd}T_{da},
\c^{\d\e_{abcd}}\big) \\
\eea
Since $\Phi_{abcd}=1_{\AG}$, $\e$ is closed 

\bea
\d\e = 0 \\
\eea
and, for each choice of $T_{ab}$, there exists an antisymmetric map 
$(m:\P'\to {\widehat{Z}})$ such that 

\bea
V_{abcd} = \exp \bigg( \frac{2i\pi}{3}\, m_{abcd} \bigg) 
\ T_{ad}T_{dc}T_{cb}T_{ba} \\
\eea
Henceforth, we will suppose that $\e=0$, i.e. that $U_{ab}$ lies 
in the connected component of $1_{\AG}$, because we want $U$ to approximate 
a continuous field. The phase of $T_{ab}$ is arbitrary and can be shifted 
freely by a gauge transformation on $m$ which, however, leaves $\Omega$ unchanged :

\bea
n:\L'& \lra & \Z_3 \\
(ab) & \lmt & n_{ab}=-n_{ba} \\
{T'}_{ab} &=& e^{\frac{2i\pi}{3}m_{ab}}\, T_{ab} \\
{m'}_{abcd} &=& m_{abcd} + (\d n)_{abcd} \\ 
\Omega' &=& e^{\frac{2i\pi}{3}\d^2 n}\, \Omega = \Omega \\
\eea
We can now express the partition function as a sum over 
$(\lambda,m)$ constrained by an integral over $T$ :

\bea
\mathcal{Z} = \sum_{\lambda m}
\int DT \ \prod_{P\in\P'} d_{\lambda_P} \widehat{v}_{\lambda_P} 
{\overline{\chi}}_{\lambda_P} \big( V_P (T,m) \big) \\
\eea
Then we use the 3-ality $\tau_\lambda\in{\widehat{Z}}\simeq\Z_3$,
defined by the restriction of $R_\lambda$ to $Z=Z(G)$, or by

\bea
\tau_\lambda=p-q\quad\mod\ 3 \\
\eea
to separate the variables $m$ and $T$ in the decomposition 
of ${\overline{\chi}}_{\lambda_P}(V_P)$ :

\bea
{\overline{\chi}}_{\lambda_P}(V_P) =
e^{\frac{2i\pi}{3}\tau_{\lambda_P}m_P}
\sum_{q_a q_b q_c q_d =1}^{d_{\lambda_P}}
\big[ R_{\lambda_P}(T_{ab}) \big]_{q_aq_b}
\big[ R_{\lambda_P}(T_{bc}) \big]_{q_bq_c}
\big[ R_{\lambda_P}(T_{cd}) \big]_{q_cq_d}
\big[ R_{\lambda_P}(T_{da}) \big]_{q_dq_a}\\
\eea
Let $\ell=(a,i)\in\L$ and let $d=a+e_i$.
Following \cite{OP}, we evaluate 
$\int dT_{\ell}$ by collecting all the factors which contain 
$T_{\ell}$ in the integrand of ${\mathcal{Z}}$. 
Let 

\bea
J_{\lambda\ell}\simeq\prod_{P\supset\ell}\{1,\cdots,d_{\lambda_P}\} \\
\eea 
index an orthonormal basis adapted to a decomposition into irreducible 
components of the tensor product of the representations $R_{\lambda_P}$
associated to the twelve oriented 2-cells $P\in\P'$ which contain $\ell$ :

\bea
\bigotimes_{P\supset\ell}R_{\lambda_P}=\bigoplus S_{\lambda_i}\\
\eea
Next, we insert the precedent decomposition into the partition function :

\bea
\mathcal{Z} &=& \sum_{\lambda m}
\bigg( 
\prod_{P\in\P'} 
d_{\lambda_P}\widehat{v}_{\lambda_P}
e^{\frac{2i\pi}{3}\tau_{\lambda_P}m_P}
\bigg)
\bigg( \prod_{\ell\in\L} K_{\lambda\ell} \bigg) \\
K_{\lambda\ell} &=&  
\sum_{qr \in J_{\lambda\ell}}
\int dT_{\ell}\ 
\bigg[
\bigotimes_{P\supset\ell}
R_{\lambda_P}
\big( T_{\ell} \big)
\bigg]_{qr} \\
\eea
Let ${\mathcal{A}}=C_{alg}(G)$ be the algebra generated by the matrix entries 
of all finite dimensional representations of $G$. We can view a representation 
$(R:G\to GL({\mathcal{V}}))$ as a vector in the $\mathcal{A}$-module

\bea
{\mathcal{M}}={\mathrm{End}}({\mathcal{V}})\otimes_{\mathbb{C}}{\mathcal{A}}\\
\eea
The integral of $R$ over $G$ provides the orthogonal projector $\mathcal{P}$
onto the $G$-invariant subspace ${\mathcal{V}}_0\subset{\mathcal{V}}$ :

\bea
\int_G R={\mathcal{P}}\in{\mathrm{End}}({\mathcal{V}})\subset{\mathcal{M}}\\
\eea
If we apply this result to $\int dT_{\ell}\,[\otimes R_{\lambda_P}]$, we obtain

\bea
{\mathcal{V}_{\lambda\ell}}&=&\bigotimes_{P\supset\ell} V_{\lambda_P} \\
{\mathcal{P}}_{\lambda\ell}&\in &{\mathrm{End}}\,({\mathcal{V}_{\lambda\ell}}) \\
K_{\lambda\ell} &=&
\sum_{qr \in J_{\lambda\ell}}
\big[ {\mathcal{P}}_{\lambda\ell} \big]_{qr} \\
\eea
Therefore, the partition function of our initial gauge fields is also 
that of a dual spin system, constrained by the projectors 
${\mathcal{P}}_{\lambda\ell}$ :

\bnea
\boxed{
\mathcal{Z} = \sum_{\lambda m}
\bigg( 
\prod_{P\in\P'} 
d_{\lambda_P}\widehat{v}_{\lambda_P}
e^{\frac{2i\pi}{3}\tau_{\lambda_P}m_P}
\bigg)
\bigg( \prod_{\ell\in\L} 
\sum_{qr \in J_{\lambda\ell}}
\big[ {\mathcal{P}}_{\lambda\ell} \big]_{qr} 
\bigg) }
\enea
\vs

\noi We can define this field theory on the dual lattice, ${C'}^\ast$, by associating 
each representation $R_{\lambda_P}$ to the dual oriented 2-cell 
$P^\ast\in {C'}^\ast$. The dual of the 2-cells which contain $\ell$ are 
the 2-cells of ${C'}^\ast$ which form the boundary of the 3-cell $\ell^\ast$ :

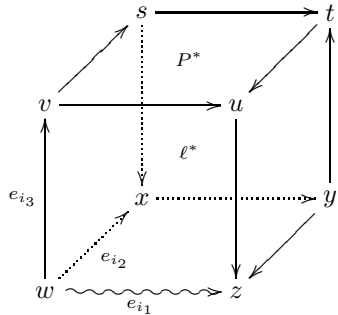
\vs
\[
\pic{
& s \ar@{}[dr]|-{P^\ast} \ar@{.>}[dd] \ar[rr] 
\ar@{}[dddr]|-{\ell^\ast} & & t \ar[dl] \\
v \ar[ur] \ar@{}[ddrr] \ar[rr] & & u \ar[dd] & \\
& x \ar@{.>}[rr] & & y \ar[uu] \ar[dl] \\
w \ar[uu]^{e_{i_3}} \ar@{.>}[ur]_{e_{i_2}} \ar@{~>}[rr]_{e_{i_1}} 
& & z & \\
}
\]
\vs

\noi In order to extend the map $(P^\ast\mapsto R_{\lambda_{P^\ast}})$
to the set of all 2-cells of the dual lattice, we must represent the action 
of $D_4$ on ${C'}^\ast$ and for this purpose, we need a dual 1-connection. 
This dual 1-connection must be a family of maps associated to the dual edge-paths and
such that the two maps associated, respectively, to the source and target of 
a dual 2-cell, $P^\ast$, can be related via the conjugation by $R_{\lambda_{P^\ast}}$. 
This relation is a dual vanishing curvature condition.

\vs
\subsection{The dual connection}

We sketch here very briefly a geometric interpretation of the dual spin system.
In the initial gauge theory, we have replaced the usual $G$-valued 1-cochain $(g_{ab})$
by the $(G\to\AG)$-valued 2-cocycle $(V_P,U_{ab})$. After a duality transformation, the theory is 
described by a field of representations $(R_{\lambda_P})$ defined on the set of oriented plaquettes of 
the dual lattice. The integration over the group variables then constrains the dual 3-curvature to vanish.
Let $\A$ denote the category of representations of $G$ over finite dimensional complex vector spaces.
The $\A$-valued 2-form $R$ is closed, i.e. for each cube $Q$ of the dual lattice, there exist
non-zero intertwiners between $\bigotimes_{P\subset Q} R_{\lambda_P}$ and the trivial representation 
$(R_0 : G \to {\mathrm{GL}}({\mathbb{C}}))$ :

\bea
\bigotimes_{P\subset Q} R_{\lambda_P} & \leftrightharpoons & R_0 \\
\eea
If Poincar\'{e}'s Lemma extends to the case of $\A$-valued $p$-forms, then there exists an 
$\A$-valued 1-form $\rho$ and a field of non-zero intertwiners between $d\rho$ and $R$ :

\bea
\rho_{wx} \otimes \rho_{xy} \otimes \rho_{yz} \otimes \rho_{zw} & \leftrightharpoons & R_{\lambda_{[wxyz]}} \\
\eea
If we integrate over such spin fields, $\rho$, then the dual Bianchi constraint is automatically satisfied.
The field $\rho$ can be viewed as a functorial connection in a fibered category with typical 
fiber $\phi_x=\A$, at each site $x$ of the dual lattice. $\rho_{xy}$ then acts by multiplication,
say on the left :

\bea
\phi_y & \lra & \phi_x \\
r & \lmt & \rho_{xy} \otimes r \\
\eea
The field $R$ then appears as a flat 2-connection on a 2-stack and $\rho$ is the 1-connection 
of a stack whose differential is the precedent 2-stack. The construction of these spaces requires 
a fully categorical approach to lattice gauge theory \cite{RO} as well as the extension of
the geometry of $G$-gerbes \cite{BM} to stacks modeled on tensor categories.

\vs
\section{Conclusions and perspectives}
\vs

In the present work, we have laid the basis of a homotopic 
approach to non-Abelian lattice gauge theories where
the field strength represents the homotopies along the 
2-cells and the gauge field transports the field strength along 
the edge-paths. A difference with the usual approach 
\cite{W,DZ,MM,GP} is in the usage of complex weights, 
$v$ and $\bar{v}$, associated to oppositely oriented
2-cells. We obtain a new closed formula for the Boltzmann weight 
of the dual spin system  which couples three fields 
of representations : one for $G$, one for its center and one 
for its group of automorphisms. We have chosen to keep the latter
vanishing in order to approximate continuous fields. 
In this dual spin model appears a projector 
${\mathcal P}_{\lambda Q}$ on each 3-cell, $Q$, of the dual lattice.
A careful definition shows that this projector
depends on the way $\p Q$ is swept. When ${\mathcal P}_{\lambda Q}\neq 0$,
any tensor product of the six representations associated to 
the six oriented plaquettes of $\p Q$ admits a non-zero 
intertwiner onto the trivial representation of $G$. 
This is the condition of vanishing 2-curvature for the dual spin model, and it is satisfied when, 
for each plaquette $P$, there exists a non-vanishing morphism of representations between $R_{\lambda_P}$ 
and the curvature $d\rho_P$ of a $\RG$-valued 1-form $(\rho_{xy})$.
The problem is then to count, for each field of highest weights $(\lambda_P)$,
the number of 1-forms $(\rho_{xy})$ for which such an intertwiner exists.
A suitable version of Poincar\'{e}'s lemma would imply that this number is non-zero 
when the 2-form $(R_{\lambda_P})$ is closed and this number would give the weight of $\rho$
in the expression of the unconstrained partition function. This is the analogue of the volume
of the gauge group in the initial theory, formulated in terms of connection and curvature.

The main lesson of the present work is that the spaces adapted to the formulation of non-Abelian duality
are fibered categories modeled on categories of $G$-torsors and $G$-modules. We must therefore reformulate 
the present homotopic approach in a purely categorical language and we hope that these methods will 
provide a notion of duality adapted to the treatment of strongly coupled gauge theories.

\vspace{1cm}

{\textbf{Acknowledgements :}} 
I thank Olivier Babelon, Marc Bellon, Robert Oeckl and Hendryk Pfeiffer 
for useful discussions at various stages of this work.

\vspace{1cm}

\begin{appendix}

\unitlength=1cm

\section{$SU_3$ and its characters}  
\vs

We recall here some standard results of group theory \cite{Ad} 
to compute the Fourier transform of $v$. 
Let $T$ be the maximal torus of $SU_3$ whose 
elements are the diagonal matrices 

\bea
D=\left( 
\begin{tabular}{ccc}
$e^{i\theta_1}$ & 0 & 0 \\ 
& & \\
0 & $e^{i(\theta_2-\theta_1)}$ & 0 \\ 
& & \\
0 & 0 & $e^{-i\theta_2}$
\end{tabular}
\right) \qquad \forall \, \theta_1 , \theta_2 \in [0,2\pi[ \\
\eea
Let $\LH$ be the Lie algebra of this torus.
Let $L_1, L_2, L_3 \in \LH^\ast$ be 
the linear forms on $\LH$ defined by

\bea
L_i (D)=D_{ii} \\
\eea
The root system of $\LH$ is the subset of $\LH^\ast$ defined by 

\bea
\Phi := \left\lbrace L_i-L_j \ : 
\ i,j \in \{ 1,2,3 \} \right\rbrace \\
\eea
The adjoint representation of $\LH$ on $\LG$ provides 
the Cartan decomposition of $\LG$ into eigenspaces of $\LH$ :

\bea
\LG = \LH \oplus 
\left( \bigoplus_{\alpha \in \Phi} \LG_\alpha \right) \\
\eea
$\LH$ corresponds to the 0 eigenvalue 
and each $\LG_\alpha$ is a one-dimensional
representation of $\LH$ :

\bea
[H,X]=\alpha(H) X \qquad \forall \, H \in \LH \ , \ 
\forall \, X \in \LG_\alpha \\
\eea
We can choose in $\Phi$ a subset $\Pi$ which is a basis of $\LH$ 
and such that every root $\alpha\in\Phi$ is a linear combination 
of elements of $\Pi$ with coefficients which are either all 
positive or all negative. This is our system of fundamental roots. 
In the present case, we take

\bea
\Pi &=& \{ \alpha_1 , \alpha_2 \} \\
\alpha_1 &=& L_1 - L_2 \\
\alpha_2 &=& L_2 - L_3 \\
\eea
$\LH$ being endowed with the scalar product induced by 
the restriction of the Killing form 

\bea
\langle H,H' \rangle = \tr (HH') 
\qquad \forall \, H,H' \in \LH \\
\eea
extended to $\LH^\ast$, we can associate to $\Pi$ 
the fundamental weights $\omega_i$ defined by 

\bea
\langle \omega_i , \alpha_j \rangle = \delta_{ij} \\
\eea
and whose expression in terms of fundamental roots is

\bea
\omega_1 = \frac{2 \alpha_1+\alpha_2}{3} 
\quad {\mathrm{and}} \quad
\omega_2 = \frac{2 \alpha_2+\alpha_1}{3} \\
\eea
The weight lattice of $SU_3$ is generated by 
$\omega_1$ and $\omega_2$ :

\bea
\Lambda_w = 
\left\lbrace \lambda = p \omega_1 + q \omega_2 \ : 
\ p,q \in \Z \right\rbrace \subset \LH^\ast \\
\eea
Each finite dimensional, ${\mathbb{C}}$-linear, irreducible
representation $(R:G\to GL(V))$ is characterized by the existence 
of a weight $\lambda \in \Lambda_w$, with positive coordinates on 
$\omega_1$ and $\omega_2$, and of a vector $u \in V$ such that 
$R(H)(u) = \lambda(H)u$ for all $H \in \LH$. The set $\widehat{G}$ 
of isomorphy classes of such representations is then parametrized 
by two positive integers :

\bnea
\boxed{
\widehat{G}\simeq\left\lbrace \lambda =
p\omega_1+q\omega_2 \ 
: \ p,q\in\N \right\rbrace }
\enea
\vs

\section{Free dynamics on the weight lattice and Fourier's inversion formula}
\vs

We give here a short proof of the fact that the Fourier transform of $w=v\bar{v}$ 
follows a free dynamics on the weight lattice of $SU_3$. 
The definition of $v$ and $w$ implies

\bea
\p_\alpha v &=& (\chi_{10}-3)\, v \\
\p_\alpha w &=& (\chi_{10}+\chi_{01}-6)\, w \\
\eea
Moreover, a straightforward computation shows that 

\bea
&& \chi_{10}\,\chi_{pq} = \chi_{p+1,q}
+\chi _{p,q-1}+\chi _{p-1,q+1} \\
&& \chi_{01}\,\chi_{pq} = \chi_{p,q+1}
+\chi_{p-1,q}+\chi_{p+1,q-1} \\
\eea
Consequently, $\widehat{v}$, $\widehat{\overline{v}}$ 
and $\widehat{w}$ satisfy the following difference/differential
equations :

\bea
\p_\alpha \widehat{v}_{pq} &=& 
\widehat{v}_{p+1,q}+\widehat{v}_{p,q-1}
+\widehat{v}_{p-1,q+1} - 3\widehat{v}_{pq} \\ 
\p_\alpha \widehat{\overline{v}}_{pq} &=& 
\widehat{\overline{v}}_{p,q+1}+\widehat{\overline{v}}_{p-1,q}
+\widehat{\overline{v}}_{p+1,q-1} - 3\widehat{\overline{v}}_{pq} \\
\p_\alpha \widehat{w}_{pq} &=&
\widehat{w}_{p+1,q}+\widehat{w}_{p,q-1}+\widehat{w}_{p-1,q+1}
+\widehat{w}_{p,q+1}+\widehat{w}_{p-1,q} +\widehat{w}_{p+1,q-1} 
- 6\widehat{w}_{pq} \\
\eea
Let $\mathcal{E}$ be the Banach space 
$\ell ^\infty (\N^2,\mathbb{C})$ of
bounded families of complex numbers indexed by $\N^2$. Let 
$\mathcal{H}$ be the Hilbert space $\ell ^2(\N^2,\mathbb{C})$ of 
square integrable families of complex numbers indexed by $\N^2$ :

\bea
\mathcal{E}=\ell^\infty (\N^2,\mathbb{C}) \quad \text{and} \quad 
\mathcal{H}=\ell^2(\N^2,\mathbb{C}) \\
\eea
$\N^2$ being endowed with the metric induced by the Killing form 
of $\LH^\ast$ via the relation (7), we define the corresponding
discrete Laplace-Beltrami 
operator, $\Delta \in B\mathcal{(}\mathcal{E})$, by 

\bea
(\Delta \phi )_{pq} := \frac{1}{6} 
\big(\phi _{p+1,q}+\phi_{p,q-1}+\phi _{p-1,q+1}+
\phi_{p,q+1} +\phi _{p-1,q}+\phi _{p+1,q-1} \big) - \phi _{pq} \\
\eea
$\Delta$ computes the mean value of a function at the 
six nearest neighbours of each site $\lambda$ and substracts 
the value at $\lambda$. In a fundamental dual Weyl chamber, 
we represent $\Delta$ by the following diagram :

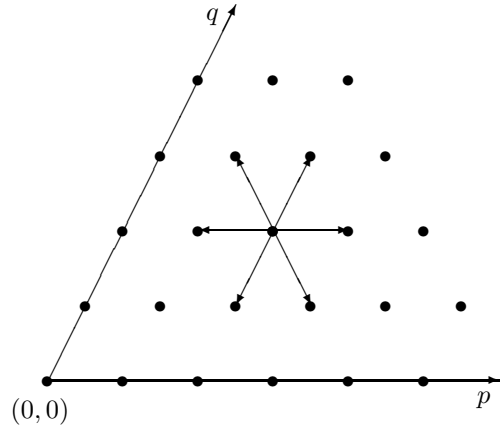
\vs
\[
\begin{picture}(3,6)(1.5,-1)
\put(-.5,-.5){$(0,0)$}
\put(0,0){\vector(1,0){6}}
\put(5.7,-0.3){$p$}
\put(0,0){\vector(1,2){2.5}}
\put(2.1,4.8){$q$}
\put(2.9,1.9){$\bullet$}
\put(3,2){\vector(1,0){1}}
\put(3,2){\vector(-1,0){1}}
\put(3,2){\vector(1,2){.5}}
\put(3,2){\vector(-1,-2){.5}}
\put(3,2){\vector(1,-2){.5}}
\put(3,2){\vector(-1,2){.5}}
\multiput(-.1,-.1)(1,0){6}{$\bullet$}
\multiput(.4,.9)(1,0){6}{$\bullet$}
\multiput(.9,1.9)(1,0){5}{$\bullet$}
\multiput(1.4,2.9)(1,0){4}{$\bullet$}
\multiput(1.9,3.9)(1,0){3}{$\bullet$}
\end{picture}
\]
\vs

\noi Thus, $\widehat{w}$ satisfies a diffusion equation on 
$\N^2\times \R_{+}$ :

\bnea
\boxed{\p_\alpha \widehat{w}= 6 \, \Delta \widehat{w}}
\enea
\vs

\noi with the boundary conditions 

\bea
\widehat{w}_{0q} = \widehat{w}_{p0}=0 \quad {\mathrm{and}} \quad 
\widehat{w}_{pq}(0) = \delta_{0p}\delta_{0q} \\
\eea
The eigenspaces of $\Delta $ in $\mathcal{E}$ are spanned by 
the eigenfunctions $(\phi_{xy})_{x,y \in \mathbb{T}}$ defined by 

\bea
\phi_{xy} \, : \,  \N^2 & \lra &\mathbb{C} \\
(p,q) & \lmt &\phi_{xy}(p,q)=x^p y^q \\
\eea
and the corresponding eigenvalues, 
$(\mu_{xy})_{x,y \in \mathbb{T}}$, are

\bea
\mu_{xy}=\frac{1}{6} \big( x+x^{-1}+y+y^{-1}+ x/y+ y/x \big) - 1 \\
\eea

\noi The spectrum, $\s (\Delta )$, of our Laplacian is the 
closure of $\mu (\mathbb{T}^2)$ in $\mathbb{R}$ : 

\bea
\s (\Delta )=\overline{\mu (\mathbb{T}^2)}
=\left[ -\frac{4}{3} \, , \, 0\right] \\
\eea
Since $\Delta $ is a bounded self-adjoint operator on $\mathcal{H}$,
we can apply the spectral theorem \cite{Z} and expand
$\widehat{v}_{pq}$ on the eigenfunctions $\phi_{xy}$ of $\Delta $ 

\bea
\widehat{v}_{pq} &=& \int_{\mathbb{T}^2} dx \, dy \, 
\langle \phi_{xy},\widehat{v} \rangle \, \phi_{xy}(p,q) \\
\langle \phi _{xy},\widehat{v} \rangle  &=& \sum_{p,q\geq 1}
\overline{\phi_{xy}(p,q)} \, \widehat{v}_{pq} \\
\phi_{xy}(p,q) &=& d_{pq} \, \chi_{pq} (M) \\
\eea
where $M={\mathrm{diag}} \, (x,y,(xy)^{-1}) \in T$, 
to obtain Fourier's inversion formula for the class functions : 

\bnea
\boxed{
v = \sum_{\lambda \in {\widehat{G}}}
{\widehat{v}}_\lambda \, d_\lambda \, {\overline{\chi}}_\lambda 
}
\enea
\vs

\section{The Fourier transform of the chiral Boltzmann weight on $SU_3$}
\vs

The characters of the irreducible, finite dimensional representations 
of $G$ can be computed by applying Weyl's character formula as
follows. For all $p,q \in \Z$, let $A_{pq}\in \mathbb{C}\,(X,Y)$ 
be the Laurent polynomial defined by

\bea
A_{pq}(X,Y)=
X^{p+q+2}Y^{q+1}+X^{-q-1}Y^{p+1}+X^{-p-1}Y^{-p-q-2}
-(X\rightleftharpoons Y) \\
\eea
$A_{pq}$ is antisymmetric under the exchange 
of $X$ and $Y$ and satisfies 

\bea
A_{-p,-q}=-A_{qp}=\overline{A_{pq}} \\
\eea
The characters of $G$ are the polynomial functions 

\bea
\chi_\lambda = \chi _{pq} = \frac{A_{pq}(L_1,L_3)}{A_{00}(L_1,L_3)} \\
\eea
One easily checks the following identities

\bea
\chi_{00} &=& 1 \\
\chi_{qp} &=& \overline{\chi_{pq}} \\
\chi _{p,-1} &=& \chi_{-1,q}=0 \\
\chi _{10}(M) &=& x+y+(xy)^{-1}=\mathrm{tr}\,(M) \\
\mathrm{Re}\,(\mathrm{tr}\,(M))&=&\frac{\chi_{01}+\chi _{10}}{2}\,(M) \\
\eea
Since $G$ is compact, and $v$ is a continuous function on 
$G$ which is invariant by conjugation, we can apply 
the Peter-Weyl theorem to approximate $v$ uniformly on $G$ 
by linear combinations of characters of $G$.
The Fourier transform of $v$ is the function 
on $\widehat{G}$ defined by

\bea
\widehat{v} \, : \,  \widehat{G} &\longrightarrow &
\bigoplus_{\lambda \in \widehat{G}}
\mathrm{End}\, (V_\lambda ) \\
\lambda  &\longmapsto &\widehat{v}_\lambda 
= \int_G v \, \pi _\lambda \\
\eea
where $R_\lambda =(V_\lambda ,\pi _\lambda )$ is a unitary
representation of $G$ for each class $\lambda \in \widehat{G}$
and $d_\lambda = \dim \, (V_\lambda)$. 
Since $v$ is a class function, Schur's lemma implies that 
$\widehat{v}_{pq} $ is a dilatation on $V_\lambda $. 
Its coefficient will still be 
written $\widehat{v}_{pq} $. Moreover, Weyl's integration formula 
expresses $\widehat{v}_{pq} $ as an integral over $T$ : 

\bea
\widehat{v}_{pq} &=& \int_G \chi _{pq} \, v \\
&=& \frac{1}{6}\int_T\left| A_{00}\right|^2\chi_{pq}\, 
\exp\left( \alpha\, \big( \chi_{10}-3 \big) \right) \\
&=& \frac{e^{-3\alpha}}{6} \int_T {\overline{A_{00}}} \, A_{pq}
\, \exp \big( \alpha \, \chi_{10} \big) \\
&=& \frac{e^{-3\alpha}}{6} \sum_{r,s \in \Z} 
C_{\lambda sr} f_{rs} (\alpha ) \\
\eea
The matrix $C_\lambda$ has a finite number of non-zero entries 
and is obtained by expanding ${\overline{A_{00}}}\,A_{pq}$ :

\bea
C_{\lambda sr} &=& \int_T {\overline{A_{00}}} \, A_{pq} 
\, X^{-r}Y^{-s} \\
\sum_{r,s \in \Z} C^\lambda_{sr} X^r Y^s &=& 
\big( X^{-2}Y^{-1}+XY^{-1}+XY^2 -Y^{-2}X^{-1}-YX^{-1}-YX^2\big) \\
&&\times\big( X^{p+q+2}Y^{q+1}+X^{-q-1}Y^{p+1}+X^{-p-1}Y^{-p-q-2} \\
&&\quad -Y^{p+q+2}X^{q+1}+Y^{-q-1}X^{p+1}+Y^{-p-1}X^{-p-q-2}\big) \\
\eea
The functions $f_{rs}(\alpha)$ are defined, for all $r,s \in \Z$, by

\bea
f_{rs} (\alpha ) &=& \int_T L_1^r L_3^s \, 
\exp \left( \alpha \, \chi_{10} \right) \\
&=& \int_0^{2\pi }\frac{d\theta _1}{2\pi }
\int_0^{2\pi }\frac{d\theta _2}{2\pi } \, 
e^{i (r \theta_1- s \theta _2 )} \, 
\exp \left( \alpha \big( e^{i\theta _1}+e^{i(\theta_2-\theta _1)} 
+ e^{-i\theta_2} \big) \right) \\
&=& \sum_{n=0}^\infty \frac{\alpha^n}{n!}
\sum_{k=0}^n 
{n \choose k}
\sum_{\ell =0}^k 
{k \choose \ell}
\int e^{i\theta_1 (k+\ell-n+r)} \int e^{ i\theta_2 (2k-\ell-n-s) } \\
&=& \sum_{n=0}^\infty \sum_{k=0}^n\sum_{\ell=0}^k
\frac{\alpha^n}{(n-k)! \, \ell ! \, (k-\ell )!} \, 
\delta _{r,n-k-\ell} \, 
\delta _{s, 2k-\ell-n} \\
\eea
In the last sum, the non-zero terms are those which satisfy 

\bea
0 \leq \ell  = \frac{n-2r-s}{3} \leq k=\frac{2n-r+s}{3} \leq n \\
\eea
Our sums being indexed by the integers for which all 
the factorials are meaningful, we have

\bea
f_{rs}(\alpha ) &=& \sum_{n \in 3 \N-r+s} \frac{ \alpha^n}
{\, \left( \frac{n+r-s}{3} \right) ! \, 
\left( \frac{n+r-s}{3} - r \right) ! 
\, \left( \frac{n+r-s}{3} + s \right) ! \, } \\
&=& \sum_{m \geq {\mathrm{max}} (0,r,-s)} 
\frac{\alpha^{3m-r+s}}{\, m! \, (m-r)! \, (m+s)! } \\
\eea
\vs
We recognize generalized hypergeometric series \cite{ASF}, 
which are defined, for $\mu ,\nu \in \N $ and 
$0 \leq \mu \leq \nu$, by

\bea
\,_\mu F_\nu \, : \, \mathbb{C}^\mu \times \mathbb{C}^\nu 
\times \mathbb{C} & \lra & \mathbb{C} \\
(a,b,z) & \lmt & \,_\mu F_\nu (a,b,z) := \sum_{n=0}^\infty 
\bigg( \frac{\, (a_1)_n (a_2)_n \cdots (a_\mu)_n \, }
{\, (b_1)_n (b_2)_n \cdots (b_\nu)_n \, } \bigg) \frac{z^n}{n!} \\
&& \qquad (a_i)_n = a_i (a_i+1) \cdots (a_i+n) \\
\eea
We thus have

\bea
f_{rs} (\alpha ) &=& 
\bigg( \frac{\alpha^{s-r}}{\, \Gamma(-r) \Gamma(s) \, }\bigg)  
\,_0 F_2 \big( \O , (-r,s) , \alpha^3 \big) 
\quad {\mathrm{if}} \quad  r<0 \quad {\mathrm{and}} \quad s>0\\
&=& \bigg( \frac{\alpha^{\vert r+s \vert + M_{rs} }}
{\, \Gamma (\vert r+s \vert ) \Gamma (M_{rs}) \, } \bigg)  
\,_0 F_2 
\big( \O , (\vert r+s \vert , M_{rs}) , \alpha^3 \big) 
\quad {\mathrm{if}} \quad  r>0 \quad {\mathrm{or}} \quad s<0 \\
\eea
where $\O$ denotes the empty family and $M_{rs}=\max (r,-s)$.
When the argument of one of the $\Gamma$ functions vanishes, 
i.e. when $r=0$ or $s=0$ or $s=-r<0$, the $\Gamma$ factor is 
replaced by 1 and the corresponding factorial coefficient in 
the series starts at the value 1.
Condensing our formulas, we obtain :

\bnea
\boxed{
\widehat{v}_\lambda = \frac{e^{-3\alpha}}{6} \, 
{\mathrm{tr}} \, \big( C_\lambda f \big) 
}
\enea
\vs

\end{appendix}


\begin{thebibliography}{99}




\bibitem[1]{Ad} 
\noi J. Frank Adams : 
\textit{Lectures on Lie Groups} ; \\
The University of Chicago Press, 1969. \\



\bibitem[2]{ASF} 
\noi G. E. Andrews, R. Askey \& R. Roy : 
\textit{Special functions} ; \\
Cambridge University Press, 2001. \\





\bibitem[3]{A} 
\noi R. Attal : 
\textit{Combinatorics of non-Abelian gerbes 
with connection and curvature} ; \\
\texttt{math-ph/0203056} ; 
to appear in {\textit{Annales de la Fondation Louis de Broglie}}. \\








\bibitem[5]{BM} 
\noi L. Breen and W. Messing :
\textit{Differential Geometry of Gerbes} ; \\
\texttt{arXiv:math.AG/0106083}. \\




\bibitem[6]{DZ} 
\noi J.-M. Drouffe and J.-B. Zuber : 
\textit{Strong coupling and mean-field methods in lattice gauge theories} ; \\
\textit{Phys. Rep.} {\textbf{102}}, 1983, pp. 1-119. \\




\bibitem[7]{FS}
\noi J. Fuchs and C. Schweigert : 
{\textit{Symmetries, Lie Algebras and Representations}} ; \\
Cambridge University Press, 1997 \\




\bibitem[8]{GP}
\noi R. Gambini and J. Pullin : 
{\textit{Loops, Knots, Gauge Theories and Quantum Gravity}} ; \\
Cambridge University Press, 1996 \\




\bibitem[9]{KW} 
\noi H. A. Kramers and G. H. Wannier :
\textit{Statistics of the two-dimensional ferromagnet} ; \\
{\textit{Phys. Rev.}} {\textbf{60}}, 1941, pp. 252-262. \\




\bibitem[10]{CWM} 
\noi S. Mac Lane : 
\textit{Categories for the Working Mathematician} ; \\
Springer-Verlag, 1997.\\



\bibitem[11]{MM} 
\noi I. Montvay and G. M\"{u}nster : 
\textit{Quantum fields on a lattice} ; \\
Cambridge University Press, 1994. \\



\bibitem[12]{RO} 
\noi R. Oeckl : 
\textit{Generalized Lattice Gauge Theory, Spin Foams and State Sum Invariants} ; \\
\texttt{hep-th/0110259}. \\



\bibitem[13]{OP} 
\noi R. Oeckl and H. Pfeiffer : 
\textit{The dual of pure non-Abelian lattice gauge theory as a spin-foam model} ; \\
{\textit{Nucl. Phys.}} {\textbf{B 598}}, No. 1-2, 2001, pp. 400-426 ; \texttt{hep-th/0008095}. \\



\bibitem[14]{OS} 
\noi K. Osterwalder and E. Seiler : 
\textit{Gauge Field Theories on a Lattice} ; \\
{\textit{Ann. Phys. }} {\textbf{110}}, 1977, pp. 440-471 ; \\




\bibitem[15]{P} 
\noi A. M. Polyakov : 
\textit{Gauge fields and strings} ; \\
Harwood Academic Publishers, 1987. \\



\bibitem[16]{Se} 
\noi J.- P. Serre : 
\textit{Repr\'{e}sentations lin\'{e}aires des groupes finis} ; \\
Hermann, 3\textsuperscript{e} \'{e}dition, 1978. \\




\bibitem[17]{W} 
\noi K. Wilson : 
\textit{Confinement of quarks} ; \\
{\textit{Phys. Rev}} {\textbf{D 10}}, 1974, pp. 2445-2459. \\




\bibitem[18]{Y} 
\noi L. G. Yaffe : 
\textit{Confinement in $SU(N)$ lattice gauge theories} ; \\
{\textit{Phys. Rev}} {\textbf{D 21}}, 1980, pp. 1574-1590. \\




\bibitem[19]{Z} 
\noi R. J. Zimmer : 
\textit{Essential Results of Functional Analysis} ; \\
The University of Chicago Press, 1990. \\



\end{thebibliography}
\end{document}